\documentclass[twocolumn]{emulateapj} 
\usepackage{amsmath}
\usepackage{amssymb}
\usepackage{amstext}
\usepackage{graphicx}

\newcommand{\be}{\begin{displaymath}}
\newcommand{\ee}{\end{displaymath}}
\newcommand{\bea}{\begin{eqnarray}}
\newcommand{\eea}{\end{eqnarray}}

\usepackage{color}
\definecolor{myColor}{rgb}{0.9,0.9,0.9}    

\begin{document}

\shorttitle{X-RAY AND OPTICAL ECLIPSES IN ULXs}
\shortauthors{POOLEY  \& RAPPAPORT}
\submitted{Submitted to ApJ on July 16, 2005}
\title{X-RAY AND OPTICAL ECLIPSES IN ULXs AS POSSIBLE INDICATORS OF BLACK HOLE MASS}

\author{D.\ Pooley\altaffilmark{1,\dag} and S.\ Rappaport\altaffilmark{2} }

\altaffiltext{1}{Astronomy Department, University of California at Berkeley, 601 Campbell Hall, Berkeley, CA 94720; {\tt dave@astron.berkeley.edu}}
\altaffiltext{2}{Department of Physics and Kavli Institute    
for Astrophysics and Space Research, MIT 37-602b, 77 Massachusetts Ave, 
                 Cambridge, MA 02139; {\tt sar@mit.edu}}
\altaffiltext{\dag}{Chandra Fellow}


\begin{abstract}

Ultraluminous X-ray sources (ULXs) with $10^{39} \lesssim L_x  <  10^{41}$ ergs s$^{-1}$
have been discovered in great numbers in external galaxies with {\em
ROSAT}, {\em Chandra}, and {\em XMM-Newton}.  The central question regarding
this important class of sources is whether they represent an extension
in the luminosity function of binary X-ray sources containing neutron
stars and stellar-mass black holes (BHs), or a new class of objects,
e.g., systems containing intermediate-mass black holes (100\,--\,1000
$M_\odot$).  We suggest searching for X-ray and optical eclipses in these systems to provide 
another diagnostic to help distinguish between these two possibilities. The sense of the effect
is that ULXs with stellar-mass black hole accretors should be at least twice as likely to exhibit eclipses as intermediate-mass black hole systems---and perhaps much more than a factor of two.
Among other system parameters, the orbital period would follow.  This would provide considerable insight as to the nature of the binary.

\end{abstract}


\section{Introduction}

The discovery and study of ultraluminous X-ray sources (ULXs) with {\em 
Einstein} (Fabbiano 1989), {\em ROSAT} (Colbert \& Ptak 2002; Roberts \&
Warwick 2000), and {\em ASCA} (Makashima et al.\ 2000) has been
greatly extended by the superior sensitivity of both {\em Chandra} and {\em XMM-Newton}  (see, e.g., 
reviews by Fabbiano \& White 2004; Colbert \& Miller 2004).  These sources 
are typically defined to be non-nuclear point sources with isotropic equivalent 
X-ray luminosities of  $L_x \ga 10^{39}$ ergs s$^{-1}$ ($2-10$
keV) and have been observed to luminosities as high as almost
$10^{41}$ ergs s$^{-1}$.  This luminosity range is a factor of $\sim 
3-300$ above the Eddington limit for neutron stars, and $\sim 1-30$ times
more than the Eddington limit for black holes of mass $\sim 10\,M_\odot$.
Thus, a key question which arises from studies of ULXs is whether the compact 
object is (1) a neutron star of mass $\sim 1.4~M_\odot$ or black hole of up to 
$\sim 15~M_\odot$ (see, e.g., Tanaka \& Lewin 1995; Greiner et al.\ 2001;
Lee et al.\ 2002; McClintock \& Remillard 2004), or (2) a black hole of
``intermediate mass'', e.g., $100-1000~M_\odot$ (e.g., Colbert \&
Mushotzky 1999).  In this work we suggest an observational diagnostic
that may help resolve this issue.

To date, over 150 ULXs have been discovered.  A recent catalog by Swartz et al.\ (2004) describes the ULX consituency of  82 galaxies observed with {\it Chandra}.  It is becoming clear that ULXs  are especially prevalent in galaxies with starburst
activity, including ones that have likely undergone a recent dynamical
encounter (e.g., Fabbiano, Zezas, \& Murray 2001; Zezas et al.\ 2002; Gao et al.\ 2003; Wolter \& Trinchieri
2003, 2004; Fabbiano \& White 2004; Colbert \&
Miller 2004).  

A number of ideas have been put forth for ways to circumvent the
problem of how $\sim 10~M_\odot$ black holes might have apparent $L_x$
values as high as $10^{41}$ ergs s$^{-1}$.  King et al.\
(2001) suggested that the radiation may be geometrically beamed so
that the true value of $L_x$ does not, in fact, exceed the Eddington
limit.  K\"ording, Falcke, \& Markoff (2002) proposed that the apparently 
super-Eddington ULXs are actually emission from microblazar jets that 
are relativistically beamed near our line of sight.   Begelman (2002) and 
Ruszkowski \& Begelman (2003) found that in radiation pressure dominated 
accretion disks super-Eddington accretion rates of a factor of $\sim$10 
can be achieved due to the existence of a photon-bubble instability in
magnetically constrained plasmas.  Thus, there may be ways in which 
stellar-mass black holes can exceed, or apparently exceed, the Eddington
limit.  Whether these models can be made to work quantitatively, remains
to be seen (see, e.g., Rappaport, Podsiadlowski, \& Pfahl 2005a).   

Whatever model is ultimately accepted for ULXs, it will have to address
the following observational facts about these sources.  
 (i) Many ULXs are found near or in star forming regions and young star clusters
 (see also Zezas et al.\ 2002;  Goad et al.\ 2002).
(ii) The X-ray spectra 
of a number of ULXs have been found to have low inner-disk
temperatures which is taken as evidence for an IMBH (Miller et al.\
2003; Miller, Fabian, \& Miller 2004a,b; Cropper et al.\ 2004). (iii) A 
substantial number of ULXs exhibit temporal variability on both 
long time scales (days--years) as well as on shorter timescales
down to $\lesssim$ hours (e.g., M74; Krauss et al.\ 2005; M51; Liu et al.\ 2002). 
Quasiperiodic oscillations in the X-ray intensity of M82 X-1 (Strohmayer
\& Mushotzky 2003) have been used to argue against beaming by 
relativistic jets in this source. (iv) Studies of the giant ionization nebulae 
surrounding a number of the ULXs (Pakull \& Mirioni 2003) seem to 
confirm the full luminosity inferred from the X-ray measurements, and 
thereby rule against relativistic beaming. (v) The optical counterpart to M81 X-1 appears to be a $\sim$20 $M_\odot$ O8 V star (Liu, Bregman, \& Seitzer 2002). The counterpart to M101 ULX-1 is consistent with a mid-B supergiant (Kuntz et al.\ 2005).  The ULX in NGC 5204 has a B0 Ib supergiant counterpart (Liu, Bregman, \& Seitzer 2004).  
Finally, we note the argument King (2004) has made that most of the 
ULXs observed in the Cartwheel galaxy cannot be IMBHs since lifetime
and formation arguments indicate that a significant fraction of the entire
mass of that galaxy would then have been involved in their production.  

In this paper we suggest that detection and observations of X-ray and/or 
optical eclipses could be very helpful in distinguishing between stellar-mass
black hole models (hereafter SMBH) and systems involving 
intermediate mass black holes (hereafter IMBH). In \S 2 we quantify the
eclipse probabilities in these two types of systems, and in \S 3 we suggest
some observational strategies for searching for eclipses.

\section{Eclipse Probabilities}

\subsection{X-Ray Eclipses}

It is an observational fact that most Roche-lobe filling (or near lobe filling) high-mass X-ray binaries exhibit X-ray eclipses while low-mass X-ray binaries do not (van Paradijs 1995).  There are two reasons for this.  The radii of the more massive donor stars occupy a larger fraction of the orbital separation than do those of the less massive donors, and this leads to a larger eclipse probability for the high-mass X-ray binaries.  Even so, the low-mass X-ray binaries exhibit fewer X-ray eclipses than can be explained by this geometric factor alone.  Milgrom (1978) proposed that accretion disks in the low-mass X-ray binary systems might have an angular opening that blocks a substantial fraction of the X-rays from the central source from impinging on the companion star, thereby rendering eclipses to external observers rarer in these systems than might otherwise be expected.  

The eclipse probability of a point X-ray source by a companion star of radius $R$ at a distance $a$, in the absence of obscuration by an accretion disk, is given by the probability that the observation angle, $\theta$, between the line of sight and  the plane of the binary orbit is less than the half angle, $\theta_\mathrm{star}$, subtended by the companion at the location of the X-ray source.  We also note that $\sin(\theta_\mathrm{star}) = R/a$.  For an assumed uniform distribution in $\theta$, the probability of observing $\theta$ less than a given value is $p(<\theta) = \sin(\theta)$.    Thus, the eclipse probability, in the absence of a disk, is
\bea
p_{\rm ecl}=R/a.
\eea

If we allow for the presence of an accretion disk that blocks X-rays for all angles $\theta \lesssim \theta_{\rm disk}$, then the eclipse probability given in eq. (1) becomes:
\bea
p_{\rm ecl} \simeq R/a - \sin(\theta_{\rm disk}) ~~~~ .
\eea
In order to evaluate these eclipse probabilities for ULXs, we assume that (i) the mass accreted by the compact object is supplied by a companion star, (ii) the donor star is filling its Roche lobe, and (iii) the accretor is a black hole.  For SMBH models the donor stars are expected to be in the range of $1 \lesssim M_\mathrm{d} \lesssim 20\,M_\odot$, while the black-hole accretors are likely to be in the range $5 \lesssim M_{\rm bh} \lesssim 15\,M_\odot$ (see, e.g., Podsiadlowski, Rappaport, \& Han 2003 and Rappaport, Podsiadlowski, \& Pfahl 2005a for detailed binary evolution models).  For IMBH models, we are not aware of any population synthesis models which     
might guide the choice of  the IMBH mass.  A number of binary evolution         
models, involving IMBH accretors, that have been made (e.g., Ivanova et al.\     
2005; Rappaport, Podsiadlowski, \& Pfahl 2005b) suggest that to obtain the      
requisite mass transfer rates for donors initiating mass transfer near the      
main sequence, donor stars with $M_\mathrm{d} \gtrsim 20\,M_\odot$ would be     
required.  Of course, donor stars initiating mass transfer later in their       
evolution would not need to be as massive. (See also the recent study by        
Blecha et al.\ 2005).  For a representative IMBH mass we simply adopted $M_{\rm bh} = 300\,M_\odot$. In all cases, we used Eggleton's (1983) analytic expression for the ratio $R_L/a$, where $R_L$ is the Roche-lobe radius (taken to be equal to the donor star radius), and where $R_L/a$ is a function of the mass ratio, $M_\mathrm{d}/M_{\rm bh}$, only.

\begin{figure}
\includegraphics[width=0.47\textwidth]{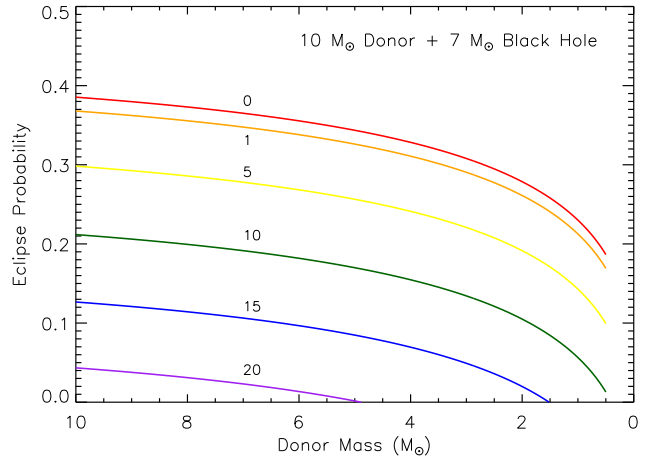}
\caption{\label{fig:SMBH}
Eclipse probabilities for an SMBH accretor of 7 $M_\odot$ as a function of donor mass.  The different curves are for a range of assumed half angular thicknesses of the accretion disk (expressed in degrees).} 
\end{figure}

The X-ray eclipse probabilities for a sequence of plausible SMBH binaries (of varying donor mass) are shown in Fig. \ref{fig:SMBH}.  Here, the maximum donor mass is $10\,M_\odot$ and the black-hole mass is fixed at $7\,M_\odot$.  Initially such a system would be subject to an interval of thermal timescale mass transfer (see Podsiadlowski et al.\ 2003), but the process would be stable and eventually the donor star would both lose most of its envelope mass and evolve up the giant branch.  These two fiducial points represent the starting and ending points of the curves.  The curves in Fig. \ref{fig:SMBH} were generated from eq. (2) with a range of assumed angular thicknesses for the accretion disk (from $\theta_{\rm disk} = 1^\circ$ to $20^\circ$).  

\begin{figure}
\includegraphics[width=0.47\textwidth]{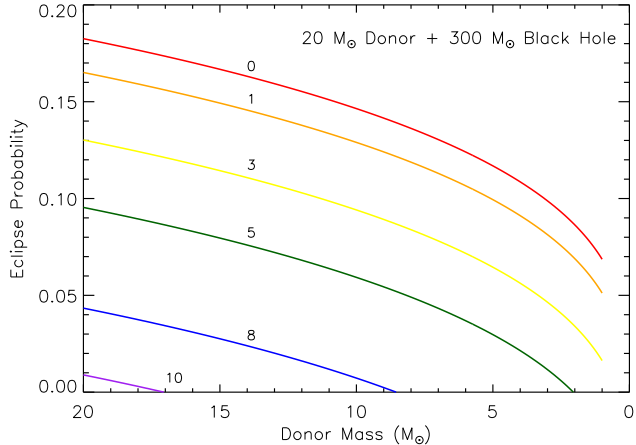}
\caption{\label{fig:IMBH}
Eclipse probabilities for an IMBH accretor of 300 $M_\odot$ as a function of donor mass.  The different curves are for a range of assumed half angular thicknesses of the accretion disk.} 
\end{figure}

As Fig. \ref{fig:SMBH} shows, the eclipse probabilities in an SMBH system should range from $\sim$40--20\% in the case of a very thin accretion disk; it would require a disk of half thickness $\sim$20$^\circ$ to nearly eliminate any probability of an eclipse.  By contrast, Fig. \ref{fig:IMBH} shows that in an IMBH system the X-ray eclipse probabilities range from about 18--7\% in the case of a very thin disk, while a disk of half thickness only $\sim$10$^\circ$ would eliminate X-ray eclipses.   

From the eclipse probability expressions and figures, we can draw four conclusions: (i) The probability of an X-ray eclipse in an SMBH ULX system can be substantial. (ii) Only for moderately thick accretion disks will the eclipses be eliminated in SMBH ULX systems. (iii) ULX systems containing IMBHs are much less likely to exhibit X-ray eclipses, even for an accretion disk with only a $\sim$5$^\circ$ half opening angular thickness.  (iv) In the event that ULX apparent luminosities are accounted for by beaming in a relativistic jet, there should be no X-ray eclipses --- assuming that the jet axis is perpendicular to the orbital plane.

\subsection{Optical Eclipses}

If there are X-ray eclipses, and even for a modest range in orbital inclination angles where there are no X-ray eclipses, there may be eclipses of the accretion disk -- and hence a significant fraction of the system optical light -- by the companion star.  Such optical eclipses would appear at the {\em same} phase as the X-ray eclipse.  The donor stars in ULXs may have masses as large as $\sim$$20\,M_\odot$ with bolometric luminosities of $\sim 3 \times 10^{38}$ ergs s$^{-1}$, but $L_{\rm bol}$ could easily be $1-2$ orders of magnitude smaller if the donor star has already transferred a substantial fraction of its envelope mass to the black-hole (see, e.g., Rappaport et al.\ 2005a).  Even when the donor star ascends the giant branch it will typically have lost most of its envelope except for $\sim 1\,M_\odot$, and its bolometric luminosity is then necessarily limited to $\sim 2 \times 10^{38}$ ergs s$^{-1}$ (i.e., the Eddington limit).

By contrast, the X-ray luminosities of the ULXs, range from $10^{39}-10^{41}$ ergs s$^{-1}$.  If only about 1\% of this energy is intercepted by the accretion disk and reprocessed into optical light (van Paradijs \& McClintock 1994; Rappaport et al.\ 2005a), then the disk may compete with, or even outshine, the companion star in the visible band.  This is especially true when one takes into account the larger bolometric correction needed for the (typically) hot companion stars compared to the cooler heated accretion disk (see Rappaport et al.\ 2005a for details).  Thus, if the visible radiation from the accretion disk is comparable to, or greater than, that from the donor star, a significant drop in optical light would be seen at the time of superior conjunction.  The details of the shape and depth of such an optical eclipse are beyond the scope of this paper, and depend on the mass and evolutionary state of the donor star, the size of the outer regions of the accretion disk where most of the optical light will originate, and, of course, the inclination angle.  

It is also possible that there would be secondary optical eclipses when the accretion disk passes in front of the donor star.  Again, the characteristics of such eclipses are beyond the scope of this paper, but also depend sensitively on the mass and evolutionary state of the donor star, the size of the accretion disk, and the orbital inclination angle. 

\begin{deluxetable}{c c c c c}
\tabletypesize{\footnotesize}
\tablewidth{0pt}
\tablecaption{Illustrative Orbital Periods and Eclipse Durations
\label{tab:gcs}}
\tablehead{
\colhead{Donor} & \colhead{Donor} &
\colhead{BH} & \colhead{$P_{\rm orb}$} & \colhead{$\tau_{\rm ecl}$} \\
\colhead{Mass ($M_\odot$)}  &
\colhead{Radius ($R_\odot$)} &
\colhead{Mass ($M_\odot$)} & 
\colhead{days} & \colhead{days} }
\startdata
10 &  4 & 7& 0.88  & 0.11 \\  
5 & 5.5 &  7 & 2.1 & 0.24  \\
2 & 19 &  7 & 22 & 2  \\
1.5 & 30 &  7 & 50 & 4  \\
1 & 55 &  7 & 150 & 11  \\
20 & 5.5 & 300 & 1.1 &  0.06  \\
10 & 6 & 300 & 1.7 &  0.08  \\
5 & 6.5 & 300 & 2.7 &  0.10  \\
2 & 24 & 300 & 31 &  0.9  \\
1 & 45 & 300 & 112 & 2.5 
\enddata
\end{deluxetable} 

\section{Observational Applications}
To identify bona fide eclipses, in either X-rays or optical light, 
good temporal coverage of the orbital period is required.  The orbital periods of the ULX binaries can be estimated from the following relation:
\bea
P_{\rm orb} \simeq 0.37 \left(\frac{R}{R_\odot}\right)^{3/2} \left(\frac{M_\mathrm{d}}{M_\odot}\right)^{-1/2} ~~ {\rm days} ~~ .
\eea
This is derived from the assumption that the donor star fills its Roche-lobe and that the Roche-lobe radius of the donor can be approximated by the relation $R/a \simeq 0.46 (M_\mathrm{d}/M_T)^{1/3}$, where $M_T$ is the total binary mass.  Strictly speaking this relation is valid only for $M_\mathrm{d} \lesssim M_{\rm bh}$,
but eq. (3) is good to $\sim$10\% accuracy even for $M_d$ up to $1.5\,M_{\rm bh}$.    Similarly, we can estimate the X-ray eclipse duration (for the case of $i \simeq 90^\circ$) as:
\bea
\tau_{\rm ecl} \simeq \frac{P_{\rm orb}}{\pi} \sin^{-1}\left\{0.46\left(\frac{M_\mathrm{d}}{M_T}\right)^{1/3}\right\}
\eea
Some illustrative values of the range of orbital periods and eclipse durations for potential SMBH and IMBH cases are given in Table 1.

Given the fact that the orbital periods for ULXs are probably in the range of $\sim 1 - 100$ days, and the eclipse durations may range from $\sim$0.1--10 days for SMBHs and $\sim$0.05--3 days for IMBH systems, there are currently very few existing X-ray observations suitable for searching for eclipses.  In general, {\it Chandra} and {\it XMM-Newton} have observed most galaxies only once or twice -- for typical durations of order a day.  Some notable exceptions are M101 (for an observation log, see Kong, Di Stefano, \& Yuan 2004) and some regions of M31 (e.g., Kong et al.\ 2003), which have numerous pointings.  However, the M31 observations are generally spaced months to years apart, making them unsuitable for searching for eclipses.  The twenty-five {\it Chandra} observations  of M101 include some sets of 2--4 observations spaced roughly days apart.  There is also a set of {\it Chandra} observations which monitor M81 every 2--4 days for a period of 6 weeks.  These observations may be promising in terms of eclipse searches.

Since {\it Chandra}, {\it XMM-Newton}, and future X-ray satellites will no doubt continue to devote time to studying the X-ray point sources and diffuse emission of nearby galaxies, the scientific utility can be maximized by considering an observational strategy that probes the expected ULX eclipse timescales.  An efficient plan would be with additional, closely-spaced observations of the Antennae and the Cartwheel (which host $\sim$20 ULXs each) as well as the very nearby galaxies with many ULXs (e.g., NGC 3034, NGC 6946, NGC 4490, and NGC 5194 which are all closer than 10 Mpc and each host five or more ULXs).

Complementary to such X-ray observations, ground- and space-based optical observations could be planned in a similar manner to sample the eclipse timescales.  The most logical candidates are those ULXs with stellar counterparts, e.g., M81 X-1, M101 ULX-1, and NGC 5204.  Obviously the identification of additional optical counterparts would be very helpful in this regard.
   
\section{Summary and Conclusions}
We have outlined a basic argument that eclipses are likely to be at least twice as probable in SMBH systems as in IMBH systems---and possibly {\em much} more probable.  We suggest that future X-ray and optical observations be designed to be sensitive to the eclipse timescales in order to characterize the eclipse properties of the ULX population.  These properties can help to determine the nature of the ULX population as whole.  Additionally, any ULX systems observed to eclipse cannot be explained by beamed emission.

The shorter period ULXs, with donor stars on or near the main-sequence, should have $P_{\rm orb} \lesssim 2.5$ days, and $\tau_{\rm ecl} \sim 1.5-6$ hr, while longer-period ULX systems, with donors on the giant branch, should have $P_{\rm orb} \gtrsim 20$ days, and $\tau_{\rm ecl} \sim 1-10$ day.  Eclipses in the former systems can readily be searched for in continuous {\em Chandra} observations of a few days (e.g., $200-300$ ksec observations).  If galaxies with $\sim$ a dozen ULXs (e.g., the Antennae) are subjected to such X-ray monitoring campaigns, then roughly 3.5 to 2.5 SMBH ULXs should exhibit eclipses (for accretion disk half thicknesses of $5^\circ-10^\circ$), compared to $\lesssim 1$ eclipse for IMBH systems with similar short orbital periods.  If the ULXs turn out to be longer orbital-period systems, then continuous Chandra searches for eclipses probably become impractical.  On the other hand, monitoring $\sim$10 ULXs in the optical from ground-based telescopes twice a night each for a couple of months should yield similar discrimination between SMBH and IMBH ULXs.  Of course the discovery of a well-defined eclipse would quickly set the stage for finding more, and the orbital period would follow.

\vspace{-.1in}

\acknowledgements
We thank Alex Filippenko and Diane Wong for helpful discussions.  One of us (SR) acknowledges support from NASA Chandra Grant NAG5-TM5-6003X.  DP gratefully acknowledges support provided by NASA through Chandra Postdoctoral Fellowship grant number PF4-50035 awarded by the Chandra X-ray Center, which is operated by the Smithsonian
Astrophysical Observatory for NASA under contract NAS8-03060.

\end{document}